\documentclass[12pt]{iopart}
\pdfoutput=1
\usepackage{graphicx}

\begin{document}

\title[Entanglement Spectra of Heisenberg Ladders of higher Spin]{Entanglement Spectra of Heisenberg Ladders of higher Spin}

\author{John Schliemann$^1$ and Andreas M. L\"auchli$^2$}

\address{$^1$Institute for Theoretical Physics, University of Regensburg,
D-93040 Regensburg, Germany\\
$^2$Institut f\"ur Theoretische Physik, 
Universit\"at Innsbruck, A-6020 Innsbruck, Austria}
\ead{john.schliemann@physik.uni-regensburg.de}
\begin{abstract}
We study the entanglement 
spectrum of Heisenberg spin ladders of arbitrary spin length $S$
in the perturbative regime of strong rung coupling. For isotropic
spin coupling the entanglement spectrum is, within first order
perturbation theory, always proportional to the energy spectrum of the
single chain with a proportionality factor being also independent of $S$. 
In particular, although the spin ladder possesses an excitation gap over
its ground state for any spin length, the entanglement spectrum is
gapless for half-integer S and gapful otherwise.   
A more complicated situation arises for anisotropic ladders of higher
spin $S\geq 1$ since here even the unperturbed ground state has a nontrivial
entanglement spectrum. Finally we discuss related issues in dimerized spin
chains.
\end{abstract}

\section{Introduction}

Quantum entanglement has by now developed to a key ingredient and tool of 
quantum many body physics\cite{Amico08,Tichy11}. More recently, the
notion of the {\em entanglement spectrum} has provided a novel
conceptual input to the field leading to new insights in the
properties of various systems\cite{Li08}. These comprise
quantum Hall monolayers at fractional filling
\cite{Li08,Regnault09,Zozulya09,Lauchli10,Thomale10a,Sterdyniak10,Thomale10b,Chandran11,Sterdyniak11,Qi11,Alba11,Sterdyniak11a,Dubail11b,Rodriguez11}, 
quantum Hall bilayers at filling factor $\nu=1$\cite{Schliemann11}, 
spin systems of one
\cite{Calabrese08,Xu08,Pollmann10a,Pollmann10b,Thomale09,Poilblanc10,Peschel11,Franchini11,Lauchli11}
and two \cite{Yao10,Cirac11,Huang11,Lou11} spatial dimensions,
and topological insulators \cite{Fidkowski10,Prodan10}. Other topics
recently covered include rotating Bose-Einstein condensates \cite{Liu11},
coupled Tomonaga-Luttinger liquids \cite{Furukawa11}, and systems
of Bose-Hubbard\cite{Deng11} and complex paired superfluids\cite{Dubail11a}.

In Ref.~\cite{Poilblanc10} Poilblanc reported the observation that chain-chain 
entanglement spectra in two-leg spin-$1/2$ ladders are remarkably similar
to the {\em energy} spectrum of a single spin-$1/2$
Heisenberg chain. Furthermore, 
the effective inverse temperature fitted to the data was found to depend 
on the ratio of the leg to the rung couplings and vanishes in the limit of 
strong rung coupling. 
Subsequently, one of the present authors observed a similarly striking
resemblance between the entanglement spectra of quantum Hall bilayers at 
$\nu=1$ and the energy 
spectrum of a single physical layer at half filling.\cite{Schliemann11}

In the case of isotropic spin-$1/2$ ladders the above observations were 
explained only shortly later by Peschel and Chung \cite{Peschel11} and by the 
present authors \cite{Lauchli11} via first-order perturbation theory around
the limit of strong rung coupling. Moreover, as shown in Ref.~\cite{Lauchli11},
an anisotropic spin-$1/2$ ladder will in general lead to an entanglement
Hamiltonian with renormalized anisotropy, and in second order perturbation
theory further corrections to the entanglement Hamiltonian occur. 

In the present paper we extend the above results for isotropic
Heisenberg ladders to the case of arbitrary spin length $S$. We
find that the entanglement spectrum is still, within first order
perturbation theory, proportional to the energy spectrum of the single chain.
This is an remarkable result since important features of the latter
spectra, such as presence or absence of an excitation gap over the ground state,
depend crucially on the spin length.\cite{Lieb61,Haldane83,Auerbach94} 
Furthermore, the 
proportionality factor (effective temperature) between the entanglement spectrum
and the single-chain energy spectrum spectrum turns out to be  
independent of the spin length.

Moreover, we also investigate $S=1$ ladders with uniaxial anisotropy. Here
we encounter the situation that already the unperturbed ground state is not
fully entangled leading to a zero-order reduced density operator which is, 
differently from the $S=1/2$ case studied earlier\cite{Lauchli11}, not
proportional to the unit matrix.

This paper is organized as follows. In section \ref{iso} 
we analyze the entanglement spectrum of isotropic Heisenberg ladders of
arbitrary spin length within first and second order perturbation theory
around the limit of strong rung coupling. In first order, the entanglement
spectrum is found to be 
proportional to the energy spectrum of the
single chain, and the  proportionality factor turns out to be
independent of $S$. Our analytical findings are compared with results
of exact numerical diagonalizations of small systems.
Section \ref{aniso} deals with spin ladders with 
anisotropic couplings and exemplifies the case of spin length $S=1$.
Technically complicated  details pertaining to sections  \ref{iso} and
\ref{aniso} are deferred to the appendices.
In section \ref{dimerchains} we remark on related observations in dimerized
spin chains of higher spin length. We close with a summary and an outlook
in section \ref{concl}.

\section{Isotropic Heisenberg ladder of arbitrary spin}
\label{iso}

We consider the Hamiltonian 
${\cal H}={\cal H}_0+{\cal H}_1$ of an isotropic spin ladder
of arbitrary spin length with
\begin{eqnarray}
{\cal H}_0 & = & J_r\sum_i\vec S_{2i}\vec S_{2i+1}\,,
\label{H0}\\
{\cal H}_1 & = & J_l\sum_i\left(\vec S_{2i}\vec S_{2i+2}
+\vec S_{2i-1}\vec S_{2i+1}\right)
\label{H1}
\end{eqnarray}
describing the coupling along the rungs and legs, respectively. The sites
on, say, the first (second) leg are denoted by even (odd) labels such that
the $i$-th rung consists of sites $2i$ and $2i+1$. All spin-$S$ operators
are taken to be dimensionless such that $J_r$, $J_l$ have dimension of
energy. The following considerations apply to finite systems with periodic
boundary conditions as well as to those of infinite size. Let us
now treat ${\cal H}_1$ as a perturbation to ${\cal H}_0$ with 
antiferromagnetic coupling, $J_r>0$. It is expected in general that the ground
state of two leg spin $S$ ladders are gapped for finite $J_r>0$ and $J_l$. For
$S=1$ for example this has been verified numerically in Ref.~\cite{Todo01}. 
The perturbative expansion around the strong rung coupling limit is therefore
an expansion into an extended phase. The unperturbed
ground state reads
\begin{equation}
|0\rangle=\bigotimes_i|s_i\rangle\,,
\label{singletprod}
\end{equation}
where the singlet state on each rung is given by, using obvious
notation,
\begin{equation}
|s_i\rangle=\sum_{m=-S}^S\frac{(-1)^{S-m}}{\sqrt{2S+1}}
|m\rangle_{2i}|-m\rangle_{2i+1}\,.
\label{rungsinglet}
\end{equation}
In order to compute the ground state in first-order perturbation theory
in ${\cal H}_1$, let us first consider the states
\begin{equation}
\vec S_{2i}|s_i\rangle=-\vec S_{2i+1}|s_i\rangle
\label{rungtriplet1}
\end{equation}
which transform under general $SU(2)$-rotations generated by 
$\vec S_{2i}+\vec S_{2i+1}$ as triplets. Thus, as a special case of the
Wigner-Eckart theorem, the states $S_{2i}^{\pm}|s_i\rangle$ and
$S_{2i}^z|s_i\rangle$ are proportional to the triplet components
$|t_i^{\pm}\rangle$ and $|t_i^z\rangle$, respectively, which are obtained
readily by expliciting Eqs.~(\ref{rungsinglet}),(\ref{rungtriplet1}) and
normalizing the results,
\begin{eqnarray}
|t_i^1\rangle & = & -\sqrt{\frac{3}{2}}
\sum_{m=-S}^{S-1}\Biggl[\frac{(-1)^{S-m}\sqrt{S(S+1)-m(m+1)}}
{\sqrt{S(S+1)(2S+1)}}\nonumber\\
 & & \qquad\qquad\times|m+1\rangle_{2i}|-m\rangle_{2i+1}\Biggr]\,,\\
|t_i^0\rangle & = & 
\sum_{m=-S}^S\frac{\sqrt{3}(-1)^{S-m}m}
{\sqrt{S(S+1)(2S+1)}}|m\rangle_{2i}|-m\rangle_{2i+1}\,,\\
|t_i^{-1}\rangle & = & -\sqrt{\frac{3}{2}}
\sum_{m=-S}^{S-1}\Biggl[\frac{(-1)^{S-m}\sqrt{S(S+1)-m(m+1)}}
{\sqrt{S(S+1)(2S+1)}}\nonumber\\
 & & \qquad\qquad\times|m\rangle_{2i}|-(m+1)\rangle_{2i+1}\Biggr]\,.
\end{eqnarray}
Considering now two neighboring rungs, the state
\begin{equation}
\vec S_{2i}\vec S_{2i+2}|s_i\rangle|s_{i+1}\rangle=
\vec S_{2i+1}\vec S_{2i+3}|s_i\rangle|s_{i+1}\rangle
\label{doublesinglet}
\end{equation}
transforms as a singlet under arbitrary rotations and, according to the
above finding, must be composed of triplet states on each rung.
In other words, this state is proportional to 
(cf. again Eq.~(\ref{rungsinglet}))
\begin{equation}
\frac{1}{\sqrt{3}}\left(|t_i^1\rangle|t_{i+1}^{-1}\rangle
-|t_i^0\rangle|t_{i+1}^0\rangle+|t_i^{-1}\rangle|t_{i+1}^1\rangle\right)\,.
\end{equation}
Indeed, a direct calculation yields the following matrix elements,
\begin{eqnarray}
\langle t_i^{\pm 1}|\langle t_{i+1}^{\mp 1}|
\vec S_{2i}\vec S_{2i+2}|s_i\rangle|s_{i+1}\rangle & = & -\frac{1}{3}S(S+1)\,,\\
\langle t_i^0|\langle t_{i+1}^0|
\vec S_{2i}\vec S_{2i+2}|s_i\rangle|s_{i+1}\rangle & = & \frac{1}{3}S(S+1)\,.
\end{eqnarray}
Note that the norm $||\vec S_{2i}\vec S_{2i+2}|s_i\rangle|s_{i+1}\rangle||^2
=(S(S+1))^2/3$ equals the sum of the squares of the above three scalar 
products, in accordance with the previous arguments.

With these results, the correction to the 
ground state within first order perturbation
theory can be formulated as
\begin{eqnarray}
 & & |1\rangle=\frac{J_l}{3J_r}S(S+1)\sum_i\Bigl[
\left(\cdots|t_i^1\rangle|t_{i+1}^{-1}\rangle
\cdots\right)\nonumber\\
 & & \qquad-
\left(\cdots|t_i^0\rangle|t_{i+1}^0\rangle
\cdots\right)+\left(\cdots|t_i^{-1}\rangle|t_{i+1}^1\rangle
\cdots\right)\Bigr]
\end{eqnarray}
where the dots denote singlet states on each rung not explicitly
specified. The reduced density operator is obtained by tracing out
one of the legs from 
$\rho^{(1)}=(|0\rangle+|1\rangle)(\langle 0|+\langle 1|)$ and is given within
first order in $J_l/J_r$ by
\begin{equation}
\rho^{(1)}_{\rm red}=\frac{1}{(2S+1)^L}\Biggl(1-\frac{2J_l}{J_r}
\sum_i\vec S_{i}\vec S_{i+1}\Biggr)
\end{equation}
with $L$ being the number of rungs.
Again within first order perturbation theory, this result can be
formulated as
\begin{equation}
\rho^{(1)}_{\rm red}=\frac{1}{Z}\exp\left(-{\cal H}^{(1)}_{\rm ent}\right)
\label{rho1iso}
\end{equation}
with $Z={\rm tr}\exp(-{\cal H}^{(1)}_{\rm ent})$ being a partition function
with respect to the entanglement Hamiltonian
\begin{equation}
{\cal H}^{(1)}_{\rm ent}=\frac{2J_l}{J_r}\sum_i\vec S_{i}\vec S_{i+1}\,.
\label{Hentiso1}
\end{equation}
Remarkably, the prefactor
\begin{equation}
\beta=\frac{2J_l}{J_r}\,,
\end{equation}
which can also be viewed as a formal inverse temperature,
is independent of the spin length $S$. 
\begin{figure}
 \includegraphics[width=\columnwidth]{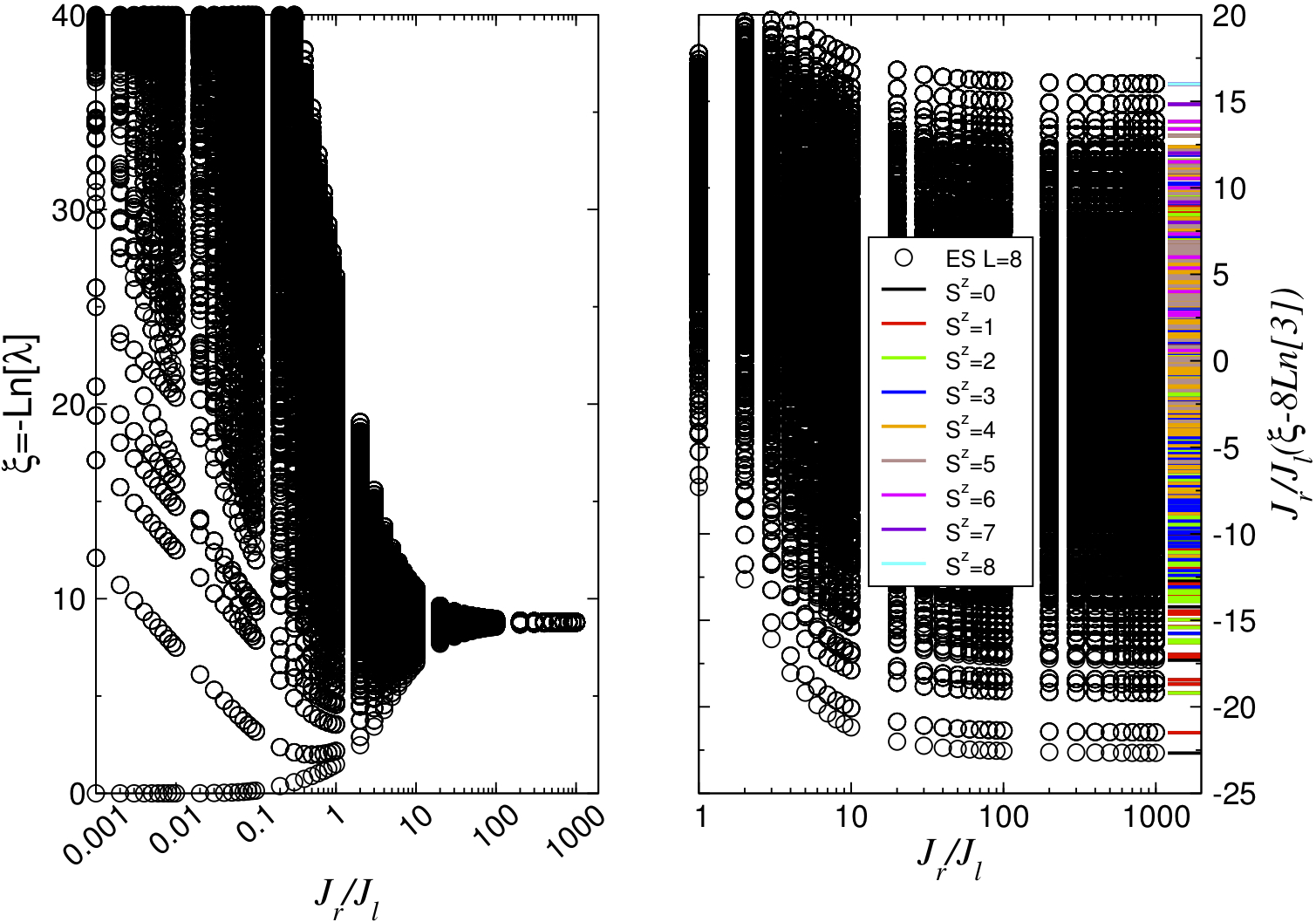}
\caption{(Color online) The entanglement spectrum of a $S=1$ spin ladder with 
$L=8$ rungs as a function of $J_r/J_l$ obtained via numerical diagonalization. 
In the right
panel we have linearly rescaled the data in order to compare it directly
with the spectrum shown on the very right 
of a single spin $S=1$ chain of eight sites labeled by
the $z$-component of the total spin. The energies have been multiplied by 2, 
according to Eq.~\eref{Hentiso1}. For practical reasons the different
sectors of total spin and linear momentum of the entanglement spectrum
are not distinguished in the plot, but these quantum nubers perfectly match
with those of the single chain.
}
\label{fig1}
\end{figure}
Fig.~\ref{fig1} shows the full entanglement spectrum of a $S=1$ spin ladder
obtained via numerical diagonalization in a wide range of $J_l/J_r$.
As seen in the right panel, in the limit of small  $J_l/J_r$
the entanglement spectrum compares 
excellently to the energy spectrum of a single Heisenberg chain, as predicted
by perturbation theory.

In summary, the entanglement spectrum within first order perturbation theory
perfectly matches the energy spectrum of a single spin chain.
This observation has already been
made in Refs.~\cite{Peschel11,Lauchli11} for $S=1/2$ and is generalized
here to arbitrary spin length. The latter result might be considered
as somewhat unexpected
since important features of energy spectra of spin chains, 
such as presence or absence of an excitation gap over the ground state,
depend crucially on the spin length.\cite{Lieb61,Haldane83,Auerbach94} 
On the other hand, the energy spectrum of the underlying spin ladder has,
in the limit of strong rung coupling, a gap over its ground state for any
spin length $S$.

The contributions of second order in $J_l/J_r$ require lengthier
calculations which are outlined in the appendix. The final result for the
reduced density in up to second order is
\begin{eqnarray}
& & \rho_{\rm red}^{(2)}=\frac{1}{(2S+1)^L}\Biggl[1-\frac{2J_l}{J_r}
\sum_i\vec S_i\vec S_{i+1}\nonumber\\
& & \qquad+\frac{1}{2}\left(\frac{2J_l}{J_r}\right)^2\Biggl[
\left(\sum_i\vec S_i\vec S_{i+1}\right)^2
+\frac{2}{5}S(S+1)\sum_i\vec S_i\vec S_{i+2}\nonumber\\
 & & \qquad\qquad\qquad-\frac{1}{10}\sum_i\left(
\left(\vec S_i\vec S_{i+1}\right)\left(\vec S_{i+1}\vec S_{i+2}\right)
+\left(\vec S_{i+1}\vec S_{i+2}\right)\left(\vec S_i\vec S_{i+1}\right)\right)
\nonumber\\
 & & \qquad\qquad\qquad-\frac{1}{6}\sum_i
\left(\left(\vec S_i\vec S_{i+1}\right)^2
+2\vec S_i\vec S_{i+1}\right)-\frac{5}{18}(S(S+1))^2L
\Biggr]\Biggr]\,,
\label{2ndrho1}
\end{eqnarray}
which can, within the same order, be reformulated as
\begin{equation}
\rho_{\rm red}^{(2)}=\frac{1}{Z}
\exp\left(-{\cal H}_{\rm ent}^{(2)}\right)
\label{2ndrho2}
\end{equation}
with $Z={\rm tr}\exp(-{\cal H}_{\rm ent}^{(2)})$ and 
\begin{eqnarray}
{\cal H}_{\rm ent}^{(2)} & = & \frac{2J_l}{J_r}\sum_i\vec S_i\vec S_{i+1}
-\left(\frac{J_l}{J_r}\right)^2\Biggl[
\frac{4}{5}S(S+1)\sum_i\vec S_i\vec S_{i+2}\nonumber\\
 & & \qquad\qquad-\frac{1}{5}\sum_i\left(
\left(\vec S_i\vec S_{i+1}\right)\left(\vec S_{i+1}\vec S_{i+2}\right)
+\left(\vec S_{i+1}\vec S_{i+2}\right)\left(\vec S_i\vec S_{i+1}\right)\right)
\nonumber\\
 & & \qquad\qquad-\frac{1}{3}\sum_i
\left(\left(\vec S_i\vec S_{i+1}\right)^2+2\vec S_i\vec S_{i+1}\right)
+\frac{1}{9}(S(S+1))^2L\Biggr]\,.
\label{2ndEH}
\end{eqnarray}
For the case $S=1/2$ the second-order entanglement Hamiltonian 
simplifies to 
\begin{equation}
{\cal H}_{\rm ent}^{(2)}=\frac{2J_l}{J_r}\sum_i\vec S_i\vec S_{i+1}
-\frac{1}{2}\left(\frac{J_l}{J_r}\right)^2
\sum_i\left(\vec S_i\vec S_{i+2}-\vec S_i\vec S_{i+1}\right)
\end{equation}
as already given in Ref.~\cite{Lauchli11}. We note that the constant term
in the Hamiltonian (\ref{2ndEH}) is strictly speaking arbitrary since
it is always canceled against contributions to the partition function. 
In Eq.~(\ref{2ndEH}) this constant was adjusted such that
${\rm tr}({\cal H}_{\rm ent}^{(2)})=0$.
\begin{figure}
 \includegraphics[width=\columnwidth]{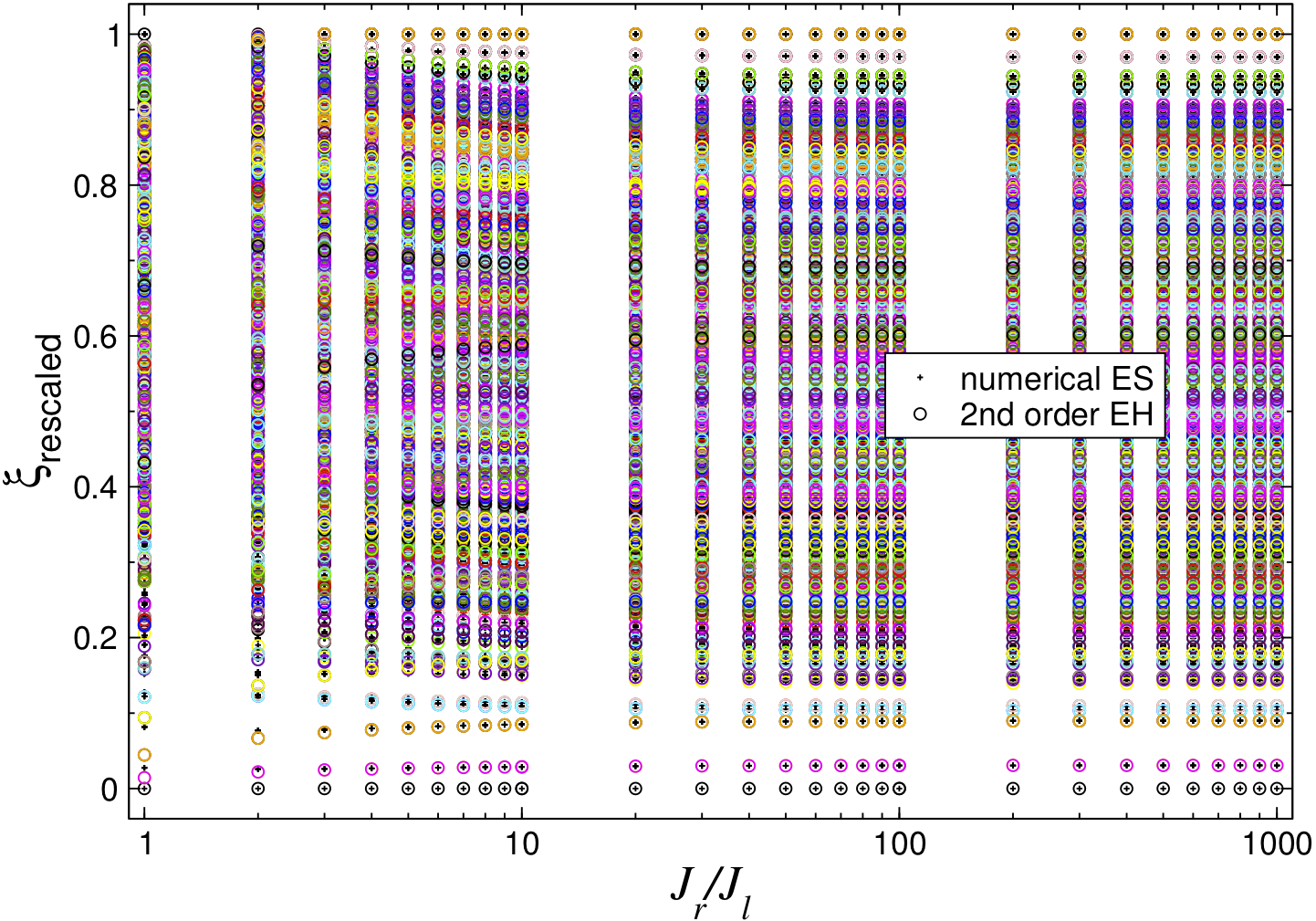}
\caption{(Color online) Cross symbols: numerical entanglement spectra obtained 
for an $S=1$ spin ladder of $L=8$ rungs. For each value of 
$J_r/J_l$ the levels have been rescaled to lie in the interval 
$[0,1]$. Circles: rescaled 
energy spectrum of the analytical second-order entanglement Hamiltonian 
(\ref{2ndEH}).
Both data show good agreement even in the vicinity of $J_r/J_l\approx 1$.
}
\label{fig2}
\end{figure}
In Fig.~\ref{fig2} we compare the numerically obtained entanglement
spectrum of an $S=1$ spin ladder with $L=8$ rungs with the energy spectrum
of the second-order entanglement Hamiltonian (\ref{2ndEH}). The plot
shows nice agreement even if the coupling strength along the legs
approaches closely the rung coupling.

\section{Ladders with anisotropic couplings}
\label{aniso}

The case of uniaxially anisotropic $S=1/2$ spin ladders was already
treated within first-order
perturbation theory in the couplings along the legs in 
Refs.~\cite{Peschel11,Lauchli11}. As an important feature, the unperturbed
ground state on each rung remains, for not too large anisotropy, the fully
entangled singlet state. This is no longer the case for larger spins since
the anisotropic contribution to the Hamiltonian couples the singlet to all
higher multiplets with even total spin which will lead to a reduction
of entanglement in the unperturbed ground state compared
to the singlet state given in Eqs.~(\ref{singletprod}),(\ref{rungsinglet}).
Indeed, it is easy to show by elementary means that for $S>1/2$
the singlet (\ref{rungsinglet}) is the only fully entangled
state of two spins $\vec S_{2i}$ and $\vec S_{2i+1}$ which lies
entirely in a single multiplet of the total spin. 

To give a specific example, let us consider the anisotropic rung Hamiltonian 
\begin{equation}
{\cal H}_0=J_r\sum_i\left(S_{2i}^xS_{2i+1}^x+S_{2i}^yS_{2i+1}^y
+\Delta S_{2i}^zS_{2i+1}^z\right)
\label{Haniso}
\end{equation}
for $S=1$. Here the ground state on rung $i$ is given by
\begin{eqnarray}
|g_i\rangle & = & \alpha_-|s_i\rangle-
\frac{\delta}{|\delta|}\alpha_+|q_i^0\rangle
\label{gs1}\\
 & = & \frac{1}{\sqrt{3}}
\Biggl(\left(\alpha_--\frac{\delta}{|\delta|}\frac{\alpha_+}{\sqrt{2}}\right)
\left[|1\rangle_{2i}|-1\rangle_{2i+1}+|-1\rangle_{2i}|1\rangle_{2i+1}\right]
\nonumber\\
 & & \qquad-\left(\alpha_-+\frac{\delta}{|\delta|}\sqrt{2}\alpha_+\right)
|0\rangle_{2i}|0\rangle_{2i+1}\Biggr)
\label{gs2}
\end{eqnarray}
with 
\begin{equation}
\alpha_{\pm}=\sqrt{\frac{1}{2}\left(1\mp
\frac{3/2-(1-\Delta)/6}{\sqrt{2+\Delta^2/4}}\right)}\,,
\end{equation}
$\delta=1-\Delta$, and the quintuplet component $|q_i^0\rangle$ being 
defined in Eq.~(\ref{quin3}).
This ground state leads to the unperturbed reduced density matrix
\begin{equation}
\rho_{\rm red}^{(0)}=\prod_i\frac{1}{3}
\left(\left(\alpha_-+\frac{\delta}{|\delta|}\sqrt{2}\alpha_+\right)^2
-3\left(\frac{\delta}{|\delta|}\sqrt{2}\alpha_-\alpha_+
+\frac{1}{2}\alpha_+^2\right)S_i^zS_i^z\right)\,,
\end{equation}
which is obviously not proportional to the unity operator but
contains for $\Delta\neq 1$ anisotropic spin couplings of arbitrary range.
In linear order in the deviation $\delta$ from isotropy we have 
$\alpha_+\approx\sqrt{2}|\delta|/9$, $\alpha_-\approx 1$, and therefore
\begin{equation}
\rho_{\rm red}^{(0)}=\frac{1}{3^L}
\left(1-\frac{2\delta}{3}\sum_i\left(S_i^zS_i^z-\frac{2}{3}\right)\right)
+{\cal O}\left(\delta^2\right)\,,
\end{equation}
or
\begin{equation}
\rho_{\rm red}^{(0)}=\frac{1}{Z}\exp\left(-{\cal H}^{(0)}_{\rm ent}\right)
+{\cal O}\left(\delta^2\right)
\end{equation}
with $Z={\rm tr}\exp(-{\cal H}^{(0)}_{\rm ent})$ and 
\begin{equation}
{\cal H}^{(0)}_{\rm ent}
=\frac{2\delta}{3}\sum_i\left(S_i^zS_i^z-\frac{2}{3}\right)
\end{equation}
Moreover, a more tedious calculation outlined in \ref{1staniso}
yields the following result for the reduced density matrix in first order
in $J_l/J_r$ and $\delta$,
\begin{eqnarray}
\rho_{\rm red}^{(1)} & = & \frac{1}{3^L}
\Biggl\{1-\frac{2\delta}{3}\sum_i\left(S_i^zS_i^z-\frac{2}{3}\right)\nonumber\\
 & & \qquad-\frac{2J_l}{J_r}\Biggl[\sum_i\left(\vec S_i\vec S_{i+1}
+\frac{\delta}{3}\left(2\left(S_i^xS_{i+1}^x+S_i^yS_{i+1}^y\right)
-S_i^zS_{i+1}^z\right)\right)\nonumber\\
& & \qquad\qquad
-\frac{\delta}{3}\Biggl(\left(\sum_i\left(S_i^zS_i^z-\frac{2}{3}\right)\right)
\left(\sum_i\vec S_i\vec S_{i+1}\right)\nonumber\\
& & \qquad\qquad\qquad
+\left(\sum_i\vec S_i\vec S_{i+1}\right)
\left(\sum_i\left(S_i^zS_i^z-\frac{2}{3}\right)\right)\Biggr)\Biggr]\Biggr\}
\label{1strhoaniso}
\end{eqnarray}
Again, the presence of an only incompletely entangled unperturbed ground
state leads to a considerably more complicated entanglement spectrum
compared to isotropic spin ladders. Let us therefore return to the case
of an isotropic Heisenberg Hamiltonian (\ref{H0}) along the rungs for again
arbitrary spin length $S$. As a perturbation we introduce the
bilinear Hamiltonian 
\begin{equation}
{\cal H}_1= J_l\sum_i\left(S_{2i}^{\alpha}A_{\alpha\beta}S_{2i+2}^{\beta}
+S_{2i-1}^{\alpha}A_{\alpha\beta}S_{2i+1}^{\beta}\right)
\label{Hleganiso}
\end{equation}
describing an arbitrarily anisotropic coupling between next neighbors
encoded, using sum convention, in the matrix $A$. Using the relations
\begin{eqnarray}
 & & S_{2i}^{\pm}|s_i\rangle=-S_{2i+1}^{\pm}|s_i\rangle
=\mp\sqrt{\frac{2}{3}S(S+1)}|t_i^{\pm 1}\rangle\,,\\
 & & S_{2i}^z|s_i\rangle=-S_{2i+1}^z|s_i\rangle
=\sqrt{\frac{1}{3}S(S+1)}|t_i^z\rangle
\end{eqnarray}
and
\begin{eqnarray}
{\rm tr}_{2i+1}|t_i^{\pm 1}\rangle\langle s_i|
 & = & \frac{\mp\sqrt{3/2}}{\sqrt{S(S+1)(2S+1)}}S_{2i}^{\pm}\,,\\
{\rm tr}_{2i+1}|t_i^z\rangle\langle s_i|
 & = & \frac{\sqrt{3}}{\sqrt{S(S+1)(2S+1)}}S_{2i}^z
\end{eqnarray}
it is easy to see that the reduced density matrix in first order in $J_l/J_r$
is of the form (\ref{rho1iso}) with
\begin{equation}
{\cal H}^{(1)}_{\rm ent}=\frac{2J_l}{J_r}
\sum_iS_{i}^{\alpha}A_{\alpha\beta}S_{i+1}^{\beta}\,.
\label{Hentaniso}
\end{equation}
Thus, the entanglement Hamiltonian in lowest nontrivial order perfectly
reproduces the functional form of the perturbation along the legs. This result 
is easily generalized to the case of bilinear couplings of longer range, or
spatially inhomogeneous couplings (with the matrix A depending on lattice
indices). The above finding (\ref{Hentaniso}) can be seen as the consequence
of conditions already formulated in Ref.~\cite{Peschel11}: (i) full
entanglement of the unperturbed ground state, and (ii) the perturbation
couples only to a single energy level (here the triplet) of the
unperturbed Hamiltonian.

\section{Dimerized spin chains}
\label{dimerchains}

The above observation that spin ladders with isotropic couplings
have perturbative 
entanglement spectra allowing for simple interpretations, can be
generalized to spin systems of other geometry. As an example, let us consider
a dimerized spin-$S$ chain (as opposed to a spin ladder) described by
${\cal H}={\cal H}_0+{\cal H}_1$ with
\begin{eqnarray}
{\cal H}_0 & = & J_0\sum_i\vec S_{2i}\vec S_{2i+1}\,,
\label{H0chain}\\
{\cal H}_1 & = & J_1\sum_i\vec S_{2i-1}\vec S_{2i}\,.
\label{H1chain}
\end{eqnarray}
For simplicity and definiteness we assume a even number $L$ of lattice sites
with periodic boundary conditions. Thus, the ground state $|G_0\rangle$
($|G_1\rangle$) of ${\cal H}_0$ (${\cal H}_1$) consists of singlets coupling
each even (odd) lattice site to the site with next higher label. Considering
now ${\cal H}_1$ as a perturbation to ${\cal H}_0$ and tracing out all,
say, odd lattice sites, one readily obtains a first-order reduced density
matrix of the form (\ref{rho1iso}) with
\begin{equation}
{\cal H}^{(1)}_{\rm ent}=\frac{J_1}{J_0}\sum_i\vec S_{2i}\vec S_{2i+2}\,.
\end{equation}
Here the missing factor of $2$ compared to Eq.~(\ref{Hentiso1}) just reflects
that we consider here a single spin chain and not two-leg ladder.

Moreover, it is also instructive to study a general linear
combination of the above two singlet states,
\begin{equation}
|\psi\rangle=\left(\cos(\theta/2)|G_0\rangle
+\sin(\theta/2)e^{i\varphi}|G_1\rangle\right)
\end{equation}
with two real parameters $\theta$, $\varphi$. For $S=1/2$ such states span the
ground state space of the Majumdar-Ghosh spin chain.\cite{Majumdar69}
Tracing out again the odd-labeled sites from $\rho=|\psi\rangle\langle\psi|$
leads to
\begin{equation}
\rho_{\rm red}=\frac{1}{(2S+1)^{L/2}}
\left(1+\frac{1}{2}\sin\theta\left(e^{i\varphi}T^++e^{-i\varphi}T\right)\right)
\end{equation}
where the operator
\begin{equation}
T=\bigotimes_{i=1}^{L/2}
\left(\sum_{m=-S}^S|m\rangle_{2i+2}\,_{2i}\langle m|\right)
\end{equation}
translates all spin states by one lattice site of the reduced chain. Thus,
the eigenstates $|k\rangle$ of $\rho_{\rm red}$ fulfill 
$T|k\rangle=e^{ik}|k\rangle$ where the wave number $k$ is restricted to
integer multiple of $4\pi/L$ by the boundary conditions. In particular,
the levels of the entanglement spectrum read as a function of $k$
\begin{equation}
\xi(k)=-\ln\left(1+\sin\theta\cos(\varphi-k)\right)
+\frac{L}{2}\ln(2S+1)\,.
\end{equation}

\section{Conclusions and outlook}
\label{concl}

We have analyzed the entanglement 
spectrum of Heisenberg spin ladders of arbitrary spin length $S$
in the perturbative regime of strong rung coupling. For isotropic
spin coupling the the entanglement spectrum turns out, within first order
perturbation theory, to be proportional to the energy spectrum of the
single chain with the proportionality factor being independent of $S$. 
Additional corrections are obtained in second order perturbation
theory which are also fully evaluated.
Due to their proportionality, the energy spectrum of a single spin chain
and the entanglement spectrum of the ladder share the property of being
gapless for half-integer spin length and gapful otherwise. This is in
contrast to the energy spectrum of the full ladder which is, for large
rung coupling, always gapped for any $S$.

A more complicated situation arises for anisotropic ladders of higher
spin $S\geq 1$ since here even the unperturbed ground state has a nontrivial
entanglement spectrum. Finally we have discussed related issues in dimerized 
spin chains.

\section*{Acknowledgements}

This work was supported by DFG via SFB631.

\appendix

\section{Second-order contribution to the ground state 
and the reduced density matrix of isotropic spin ladders}
\label{2nd}

Defining the two-rung singlet state
\begin{equation}
\sigma^{(1)}_{ij}=\frac{1}{\sqrt{3}}\left(|t_i^1\rangle|t_j^{-1}\rangle
-|t_i^0\rangle|t_j^0\rangle+|t_i^{-1}\rangle|t_j^1\rangle\right)
\end{equation}
the first-order contribution can be formulated as
\begin{equation}
|1\rangle=\frac{J_l}{\sqrt{3}J_r}S(S+1)\sum_i
\left(\cdots\sigma^{(1)}_{i,i+1}\cdots\right)\,.
\end{equation}
The second-order correction to a non-degenerate state $|0\rangle$ is given
by the general expression (using obvious notation)
\begin{eqnarray}
|2\rangle & = & 
\sum_{\alpha\neq 0}
|\alpha\rangle\frac{\langle 0|{\cal H}_1|0\rangle
\langle\alpha|{\cal H}_1|0\rangle}
{(E_0-E_{\alpha})^2}-\frac{1}{2}|0\rangle\sum_{\alpha\neq 0}
\frac{\langle 0|{\cal H}_1|\alpha\rangle
\langle\alpha|{\cal H}_1|0\rangle}
{(E_0-E_{\alpha})^2}\nonumber\\
 & + & \sum_{\alpha\neq 0\neq\beta}
|\alpha\rangle\frac{\langle\alpha|{\cal H}_1|\beta\rangle
\langle\beta|{\cal H}_1|0\rangle}
{(E_0-E_{\alpha})(E_0-E_{\beta})}\,.
\label{2ndgen}
\end{eqnarray}
The first term of the above r.h.s. does not contribute in the present case,
and in the second term again only triplet excitations on the rungs occur which
can be evaluated using the matrix elements calculated previously. The last
contribution is more complicated since here (for $S>1/2$)
also quintuplet excitations on a given rung
come into play. The components of such a quintuplet on rung $i$ read explicitly
\begin{eqnarray}
|q_i^2\rangle & = & Q(S)
\sum_{m=-S+1}^{S-1}\Bigl[(-1)^{S-m}\sqrt{S(S+1)-m(m-1)}\nonumber\\
 & & \qquad\qquad\cdot
\sqrt{S(S+1)-m(m+1)}|m+1\rangle_{2i}|-m+1\rangle_{2i+1}\Bigr]\,,
\label{quin1}\\
|q_i^1\rangle & = & Q(S)
\sum_{m=-S}^{S-1}\Bigl[(-1)^{S-m}\sqrt{S(S+1)-m(m+1)}\nonumber\\
 & & \qquad\qquad\cdot(2m+1)|m+1\rangle_{2i}|-m\rangle_{2i+1}\Bigr]\,,
\label{quin2}\\
|q_i^0\rangle & = & Q(S)\sqrt{\frac{2}{3}}
\sum_{m=-S}^{S}\left[(-1)^{S-m}\left(S(S+1)-3m^2\right)
|m\rangle_{2i}|-m\rangle_{2i+1}\right]\,,
\label{quin3}\\
|q_i^{-1}\rangle & = & Q(S)
\sum_{m=-S}^{S-1}\Bigl[(-1)^{S-m}\sqrt{S(S+1)-m(m+1)}\nonumber\\
 & & \qquad\qquad\cdot(2m+1)|m\rangle_{2i}|-m-1\rangle_{2i+1}\Bigr]\,,
\label{quin4}\\
|q_i^{-2}\rangle & = & Q(S)
\sum_{m=-S+1}^{S-1}\Bigl[(-1)^{S-m}\sqrt{S(S+1)-m(m-1)}\nonumber\\
 & & \qquad\qquad\cdot
\sqrt{S(S+1)-m(m+1)}|m-1\rangle_{2i}|-m-1\rangle_{2i+1}\Bigr]\,,
\label{quin5}
\end{eqnarray}
where
\begin{equation}
Q(S)=\frac{-\sqrt{15/2}}{\sqrt{S(S+1)(2S+1)(2S+3)(2S-1)}}
\end{equation}
The highest-weight state $|q_i^2\rangle$ of the above multiplet is again
readily obtained by, e.g., applying $S_{2i}^+$ on $|t_i^1\rangle$ and 
normalizing the result. Now in order to compute the second-order correction
to the ground state one needs to evaluate expressions of the form
\begin{equation}
\left(\vec S_{2i}\vec S_{2i+2}+\vec S_{2i+1}\vec S_{2i+3}\right)
\left(\cdots\sigma^{(1)}_{j,j+1}\cdots\right)\,.
\end{equation}
If the rungs involved do not overlap, i.e. $|i-j|>1$, we just obtain 
another independent singlet $\sigma^{(1)}_{i,i+1}$ composed of triplets
on rungs $i$ and $i+1$. In the remaining cases one finds, after somewhat
tedious calculations, the following expansions,
\begin{eqnarray}
& & \left(\vec S_{2i-2}\vec S_{2i}+\vec S_{2i-1}\vec S_{2i+1}\right)
\left(\cdots\sigma^{(1)}_{i,i+1}\cdots\right)\nonumber\\
& & \qquad=-\frac{\sqrt{2}}{3}\sqrt{S(S+1)(2S+3)(2S-1)}
\left(\cdots\sigma_{i-1,i,i+1}\cdots\right)\nonumber\\
& & \qquad\qquad
+\frac{2}{3}S(S+1)\left(\cdots\sigma^{(1)}_{i-1,i+1}\cdots\right)\,,\\
& & \left(\vec S_{2i}\vec S_{2i+2}+\vec S_{2i+1}\vec S_{2i+3}\right)
\left(\cdots\sigma^{(1)}_{i,i+1}\cdots\right)\nonumber\\
& & \qquad=-\frac{1}{\sqrt{15}}(2S+3)(2S-1)
\left(\cdots\sigma^{(2)}_{i,i+1}\cdots\right)\nonumber\\
& & \qquad\qquad-\left(\cdots\sigma^{(1)}_{i,i+1}\cdots\right)
-\frac{2}{\sqrt{3}}S(S+1)|0\rangle\,.
\end{eqnarray}
Here we have introduced two further types of singlet states,
\begin{eqnarray}
\sigma^{(2)}_{ij} & = &\frac{1}{\sqrt{5}}\Bigl(|q_i^2\rangle|q_j^{-2}\rangle
-|q_i^1\rangle|q_j^{-1}\rangle+|q_i^0\rangle|q_j^0\rangle\nonumber\\
& & \qquad\qquad-|q_i^{-1}\rangle|q_j^1\rangle+|q_i^{-2}\rangle|q_j^2\rangle
\Bigr)\,,\\
\sigma_{ijk} & = & \frac{1}{\sqrt{3}}
\Bigl(|\tau_{ij}^1\rangle|t_k^{-1}\rangle
-|\tau_{ij}^0\rangle|t_k^0\rangle+|\tau_{ij}^{-1}\rangle|t_k^1\rangle\Bigr)\\
 & = & \frac{1}{\sqrt{3}}
\Bigl(|t_i^1\rangle|\tau_{kj}^{-1}\rangle-|t_i^0\rangle|\tau_{kj}^0\rangle
+|t_i^{-1}\rangle|\tau_{kj}^1\rangle\Bigr)\,,
\end{eqnarray}
where
\begin{eqnarray}
|\tau_{ij}^1\rangle & = & \sqrt{\frac{1}{10}}|t_i^1\rangle|q_j^0\rangle
-\sqrt{\frac{3}{10}}|t_i^0\rangle|q_j^1\rangle
+\sqrt{\frac{3}{5}}|t_i^{-1}\rangle|q_j^2\rangle\\
|\tau_{ij}^0\rangle & = & \sqrt{\frac{3}{10}}|t_i^1\rangle|q_j^{-1}\rangle
-\sqrt{\frac{2}{5}}|t_i^0\rangle|q_j^0\rangle
+\sqrt{\frac{3}{10}}|t_i^{-1}\rangle|q_j^1\rangle\\
|\tau_{ij}^{-1}\rangle & = & \sqrt{\frac{3}{5}}|t_i^1\rangle|q_j^{-2}\rangle
-\sqrt{\frac{3}{10}}|t_i^0\rangle|q_j^{-1}\rangle
+\sqrt{\frac{1}{10}}|t_i^{-1}\rangle|q_j^0\rangle
\end{eqnarray}
are the components of a two-rung triplet state composed of a triplet on
rung $i$ and a quintuplet on rung $j$. Summing up, the second-order correction
to the ground state can be written as
\begin{eqnarray}
|2\rangle & = & -\left(\frac{J_l}{J_r}\right)^2\frac{(S(S+1))^2}{6}L|0\rangle
\nonumber\\
 & + & \left(\frac{J_l}{J_r}\right)^2\Biggl[\frac{(S(S+1))^2}{6}
\sum_{|i-j|>1}\left(\cdots\sigma^{(1)}_{i,i+1}\cdots
\sigma^{(1)}_{j,j+1}\cdots\right)\nonumber\\
 & & \qquad\quad+\frac{2\sqrt{2}}{15\sqrt{3}}(S(S+1))^{3/2}
\sqrt{(2S+3)(2S-1)}\sum_i\left(\cdots\sigma_{i-1,i,i+1}\cdots\right)\nonumber\\
 & & \qquad\quad-\frac{2(S(S+1))^2}{3\sqrt{3}}
\sum_i\left(\cdots\sigma^{(1)}_{i,i+2}\cdots\right)\nonumber\\
 & & \qquad\quad+\frac{1}{18\sqrt{5}}S(S+1)(2S+3)(2S-1)
\sum_i\left(\cdots\sigma^{(2)}_{i,i+1}\cdots\right)\nonumber\\
 & & \qquad\quad+\frac{1}{2\sqrt{3}}
\sum_i\left(\cdots\sigma^{(1)}_{i,i+1}\cdots\right)\Biggr]\,,
\label{2ndstate}
\end{eqnarray}
where the first line on the r.h.s results from the second term in
Eq.~(\ref{2ndgen}), and all other contributions stem from the
second line there. With these results at hand, the second-order
correction to the reduced density matrix is given by the sum of the
expressions
\begin{eqnarray}
  & & {\rm tr}_{\rm 1leg}\left(|1\rangle\langle 1|\right)=
\left(\frac{J_l}{J_r}\right)^2\frac{1}{(2S+1)^L}
\left(\sum_i\vec S_i\vec S_{i+1}\right)^2\,,\\
 & & {\rm tr}_{\rm 1leg}\left(|0\rangle\langle 2|+|2\rangle\langle 0|\right)
=\left(\frac{J_l}{J_r}\right)^2\frac{1}{(2S+1)^L}
\Biggl[\left(\sum_i\vec S_i\vec S_{i+1}\right)^2\nonumber\\
 & & \qquad\qquad\qquad-\frac{1}{5}\sum_i\left(
\left(\vec S_{i}\vec S_{i+1}\right)\left(\vec S_{i+1}\vec S_{i+2}\right)
+\left(\vec S_{i+1}\vec S_{i+2}\right)\left(\vec S_{i}\vec S_{i+1}\right)\right)
\nonumber\\
 & & \qquad\qquad\qquad-\frac{1}{3}\sum_i
\left(\left(\vec S_{i}\vec S_{i+1}\right)^2
+2\vec S_{i}\vec S_{i+1}\right)\nonumber\\
 & & \qquad\qquad\qquad+\frac{4}{5}S(S+1)\sum_i\vec S_{i}\vec S_{i+2}
-\frac{5}{9}(S(S+1))^2L\Biggl]\,,
\end{eqnarray}
which lead to Eq.~(\ref{2ndrho1}). The above considerations refer explicitly
to the case $S>1/2$ due to the presence of the quintuplets
which are absent for $S=1/2$. In the second-order
correction to the ground state  rungs with such
states occur in the second and the fourth line of Eq.~(\ref{2ndstate}). 
The corresponding contributions to the reduced density matrix read
\begin{eqnarray}
 & & \left(\frac{J_l}{J_r}\right)^2\frac{4/5}{(2S+1)^L}
\Biggl[\sum_i\left(
\left(\vec S_{i}\vec S_{i+1}\right)\left(\vec S_{i+1}\vec S_{i+2}\right)
+\left(\vec S_{i+1}\vec S_{i+2}\right)\left(\vec S_{i}\vec S_{i+1}\right)\right)
\nonumber\\
 & & \qquad\qquad-\frac{2}{3}S(S+1)\sum_i\vec S_{i}\vec S_{i+2}\Biggr]\,,\\
& & \left(\frac{J_l}{J_r}\right)^2\frac{2/3}{(2S+1)^L}
\sum_i\left(\left(\vec S_i\vec S_{i+1}\right)^2
+\frac{1}{2}\vec S_i\vec S_{i+1}-\frac{1}{3}(S(S+1))^2\right)\,,
\end{eqnarray}
respectively, and vanish for $S=1/2$ due to elementary identities of
Pauli matrices. As a result, the $S=1/2$ reduced density matrix in up to
second order in $J_l/J_r$ is given by
\begin{eqnarray}
\rho_{\rm red}^{(2)} & = & \frac{1}{2^L}\Biggl[1-\frac{2J_l}{J_r}
\sum_i\vec S_i\vec S_{i+1}+\frac{1}{2}\left(\frac{2J_l}{J_r}\right)^2\Biggl[
\left(\sum_i\vec S_i\vec S_{i+1}\right)^2-\frac{3}{16}L\nonumber\\
 & & \qquad\qquad
+\frac{1}{4}\sum_i\left(\vec S_i\vec S_{i+2}-\vec S_i\vec S_{i+1}\right)
\Biggr]\Biggr]\,,
\end{eqnarray}
which was already found in Ref.~\cite{Lauchli11}.

\section{First-order perturbation theory for anisotropic ladders}
\label{1staniso}

We now consider an Hamiltonian of the form (\ref{Haniso}) acting with
amplitude $J_l$ (cf. Eq.~(\ref{H1})) as a perturbation along the legs. 
Its action on the each rung pair is given by
\begin{eqnarray}
 & & \Bigl(S_{2i}^xS_{2i+2}^x+S_{2i}^yS_{2i+2}^y
+\Delta S_{2i}^zS_{2i+2}^z\nonumber\\
& & \quad+S_{2i+1}^xS_{2i+3}^x+S_{2i+1}^yS_{2i+3}^y
+\Delta S_{2i+1}^zS_{2i+3}^z\Bigr)|g_i\rangle|g_{i+1}\rangle\nonumber\\
 & & =\frac{4}{3}\Delta^2\left(\alpha_-
-\frac{\delta}{|\delta|}\frac{\alpha_+}{\sqrt{2}}\right)^2
|t_i^0\rangle|t_{i+1}^0\rangle\nonumber\\
 & & \quad-\frac{1}{3}\left(2\alpha_-
+\frac{\delta}{|\delta|}\frac{\alpha_+}{\sqrt{2}}\right)^2
\left(|t_i^1\rangle|t_{i+1}^{-1}\rangle
+|t_i^{-1}\rangle|t_{i+1}^1\rangle\right)\nonumber\\
 & & \quad+\frac{3}{2}\alpha_+^2
\left(|q_i^1\rangle|q_{i+1}^{-1}\rangle
+|q_i^{-1}\rangle|q_{i+1}^1\rangle\right)\,.
\end{eqnarray}
Remarkably, no coupling to the excited rung eigenstates 
\begin{equation}
|e_i\rangle=\alpha_+|s_i\rangle+\frac{\delta}{|\delta|}\alpha_-|q_i^0\rangle
\end{equation}
occurs, which are the other orthogonal states spanned by
$|s_i\rangle$ and $|q_i^0\rangle$. Now the first-order correction to the
ground state is readily obtained as
\begin{eqnarray}
|1\rangle & = & \frac{J_l}{J_r}\frac{2\Delta^2/3}{E_g+2-\Delta}
\left(\alpha_--\frac{\delta}{|\delta|}\frac{\alpha_+}{\sqrt{2}}\right)^2
\sum_i\left(\cdots|t_i^0\rangle|t_{i+1}^0\rangle\cdots\right)\nonumber\\
 & - & \frac{J_l}{J_r}\frac{1/6}{E_g+1}
\left(2\alpha_-+\frac{\delta}{|\delta|}\frac{\alpha_+}{\sqrt{2}}\right)^2
\sum_i\left(\cdots\left[|t_i^1\rangle|t_{i+1}^{-1}\rangle
+|t_i^{-1}\rangle|t_{i+1}^1\rangle\right]\cdots\right)\nonumber\\
& + & \frac{J_l}{J_r}\frac{3/4}{E_g-1}\alpha_+^2
\sum_i\left(\cdots\left[|q_i^1\rangle|q_{i+1}^{-1}\rangle
+|q_i^{-1}\rangle|q_{i+1}^1\rangle\right]\cdots\right)
\end{eqnarray}
where the dots denote states $|g_j\rangle$ (cf. Eq.~(\ref{gs1}))
on all rungs $j$ not explicitly specified, and
\begin{equation}
E_g=-\frac{1}{2}+\frac{1-\Delta}{2}-\sqrt{2+\frac{\Delta^2}{4}}
\end{equation}
such that the ground state energy on each unperturbed rung is $J_rE_g$.
With these results at hand it is straightforward to compute the reduced
density matrix within first order in $J_l/J_r$. For general anisotropy,
however, the resulting expressions are rather cumbersome and difficult
to interpret. We therefore concentrate here on the linear order in 
$\delta=1-\Delta$,
 \begin{eqnarray}
|1\rangle & = & -\frac{J_l}{J_r}\frac{2}{3}
\left(1-\frac{5}{9}\delta\right)
\sum_i\left(\cdots|t_i^0\rangle|t_{i+1}^0\rangle\cdots\right)\nonumber\\
 & + & \frac{J_l}{J_r}\frac{2}{3}
\left(1+\frac{7}{9}\delta\right)
\sum_i\left(\cdots\left[|t_i^1\rangle|t_{i+1}^{-1}\rangle
+|t_i^{-1}\rangle|t_{i+1}^1\rangle\right]\cdots\right)
+{\cal O}\left(\delta^2\right)
\end{eqnarray}
where on each unspecified rung we have states of the form
\begin{equation}
|g_i\rangle=|s_i\rangle-
\frac{\sqrt{2}\delta}{9}|q_i^0\rangle
+{\cal O}\left(\delta^2\right)
\end{equation}
which also build up the unperturbed ground state. Note that in linear order
in $\delta$ no contribution to $|1\rangle$ from quintuplets occurs.
Finally, tracing out one of the legs leads one finds for the reduced 
density matrix
\begin{eqnarray}
\rho_{\rm red}^{(1)} & = & \frac{1}{3^L}
\Biggl\{1-\frac{2\delta}{3}\sum_i\left(S_i^zS_i^z-\frac{2}{3}\right)\nonumber\\
 & & \qquad-\frac{2J_l}{J_r}\sum_i\Biggl[\vec S_i\vec S_{i+1}
\left(1-\frac{2\delta}{3}\sum_{j\neq i,i+1}
\left(S_j^zS_j^z-\frac{2}{3}\right)\right)\nonumber\\
 & & \qquad\qquad +\frac{8\delta}{9}
\left(S_i^xS_{i+1}^x+S_i^yS_{i+1}^y\right)
-\frac{7\delta}{9}S_i^zS_{i+1}^z\Biggr]\Biggl\}\nonumber\\
 & & +{\cal O}\left(\delta^2\right)\,.
\end{eqnarray}
Using the identities $S^zS^zS^{\pm}+S^{\pm}S^zS^z=S^{\pm}$ and
$(S^z)^3=S^z$ for $S=1$ operators leads to the more symmetric
form (\ref{1strhoaniso}).
{}

\end{document}